\documentstyle[pra,aps,twocolumn,floats,epsfig]{revtex}

\begin{document}

\draft

\wideabs{

\title{Optical shielding of cold collisions in blue-detuned near-resonant
optical lattices}

\author{J. Piilo$^1$ and K.-A. Suominen$^{1,2}$}

\address{$^1$Helsinki Institute of Physics, PL 64, FIN-00014 Helsingin
yliopisto,
Finland}

\address{$^2$Department of Physics, University of Turku,
FIN-20014 Turun yliopisto, Finland}

\date{\today}

\maketitle

\begin{abstract}
We report Monte Carlo wave function simulation results for two colliding atoms
in a blue-detuned near-resonant $J=1\rightarrow J=1$ optical lattice.  Our
results show that complete optical shielding of collisions can be achieved
within the lattice with suitably selected and realistic laser field
parameters. More importantly, our results demonstrate that the shielding
effect does not interfere with the actual trapping and cooling process, and it
is produced by the lattice lasers themselves, without the need to use
additional laser beams.
\end{abstract}

}

\narrowtext

\section{Introduction}

In laser cooling and trapping of neutral atoms spatial variations in the
polarization and intensity of light can be effectively turned into potentials
for atoms~\cite{vanderStraten99}. Optical lattices are an important
application of this technique~\cite{Jessen96}. By using suitable laser field
configurations and parameters one can achieve localization of atoms into
lattice sites, at least for some period of time on average. In practice there
are two distinct cases of optical lattices. For near-resonant light the lasers
are responsible for both cooling and trapping of atoms, usually achieved by a
strong scattering of photons (energy dissipation). The basic example is the
Sisyphus or polarization gradient cooling mechanism~\cite{Dalibard89}. In
contrast, with far off-resonant light we obtain only a conservative lattice
potential without the cooling mechanism, with a strong reduction in photon
scattering. This allows many delicate experiments involving quantum
coherences, such as Bose-Einstein condensation
~\cite{Hensinger01,Greiner02,Pedri01,Morsch01}.

At suitably large atomic densities, about $10^{12}$ atoms/cm$^3$, the
atom-atom interactions are expected to become important in laser
cooling and trapping~\cite{Holland94,Weiner99}. For far off-resonant lattices
their role can be easily controlled, and one can even consider studying
controlled collisions~\cite{Jaksch98}, or using the atom-atom interactions to
study quantum computing and entanglement~\cite{Brennen99,Jaksch99}. Recently it
has become possible to make densely occupied far off--resonant
lattices~\cite{DePue99}. For near-resonant optical lattices the situation is
more complicated, because the main atom-atom effects are light-assisted
inelastic collisions, which interfere with the laser cooling process. Another
aspect is that although strong photon scattering leads typically to efficient
dissipation and cooling, in suitably dense atomic gases the reabsorption of
scattered photons causes also unwanted heating of the atomic cloud.
Experimentally the case of densely populated near-resonant lattices has not
been studied very much so far, as high occupation densities are hard to
produce due to the above-mentioned problems. These problems are present also
in basic magneto-optical traps (MOT) so one does not get much gain in density
by using a MOT as a preliminary stage.

For near-resonant laser light we have two situations depending whether the
laser frequency is slightly above the atomic transition (blue-detuned light)
or slightly below it (red-detuned light). In the more commonly applied
red-detuned case one obtains the lattice structure rather easily, and can go
below the Doppler limit for laser cooling due to the Sisyphus cooling
mechanism. For densely populated lattices this means, however, the
above-mentioned appearance of strongly inelastic collisional radiative
processes that cause heating and loss of atoms~\cite{Weiner99}.

In this paper we look at the case of near-resonant blue-detuned lattice, which
is different from the red-detuned case. It is not straightforward to construct
a blue-detuned lattice with an efficient cooling mechanism. There are,
however, schemes for doing this, based on applying an external magnetic
field~\cite{Grynberg94,Hemmerich95}, and we have adapted one of them as the
basis of our studies. The important benefit of using a near-resonant
blue-detuned lattice is that the photon scattering is reduced strongly
compared to the red-detuned case. This diminishes the role of photon
reabsorption as a heating mechanism, and thus increases the relevance of the
light-assisted collisions in the complicated thermodynamics of the atomic
cloud.

With blue-detuned light the nature of the light-assisted collisions is also
changed, and instead of strong inelastic processes we have optical shielding
of collisions~\cite{Weiner99}. For weak fields this mechanism is
incomplete and can lead to some heating of atoms, but barely to any strong
loss. For stronger fields the inelastic contribution disappears, also in the
sense that optical shielding forbids close ground-ground encounters which can
lead to atom loss or heating, in addition to the light-assisted processes.
Optical shielding has been studied both
experimentally~\cite{Bali94,Marcassa94,Katori94,Suominen96b,Sukenik98} and
theoretically~\cite{Suominen95,Yurovsky97,Napolitano97} with laser cooled
atoms in magneto-optical traps, but not in optical lattices to our knowledge.
None of the theoretical studies so far have treated the cooling process and
the shielding process simultaneously.

Thus we expect that, {\it a priori}, it is better to use blue-detuned
lattices for efficient cooling while aiming at an increase in the atomic
density in lattices. Matters are, however, complicated by the presence of
the Zeeman states with different energies. Also, the mixing of cooling
and collisions may produce unexpected effects if the two processes are not
separable. Finally, there are indications in previous MOT studies that at
very strong fields off-resonant processes can become important. Thus, by using
the technique of our earlier study~\cite{Piilo01a,Piilo01b}, we have examined
the dynamics of two colliding atoms in a one-dimensional blue-detuned optical
lattice. It turns out that all the above-mentioned aspects, incomplete and
complete shielding, and the off-resonant processes, are present also for the
blue-detuned lattices. But more importantly, it turns out that the shielding
mechanism and the cooling and trapping mechanism can coexist without
interfering with each other.

This paper is constructed as follows. First we briefly review in
Sec.~\ref{sec:bluelatt} the basic properties and construction of
blue-detuned optical lattices, and in Sec.~\ref{sec:Shielding} the optical
shielding process in collisions. In Sec.~\ref{sec:Frame} we present how
to treat the two situations within the same dynamical model with Monte
Carlo wave function (MCWF) simulations. The results are given in
Sec.~\ref{sec:Results} and finally discussed in
Sec.~\ref{sec:Conclusions}.

\section{Blue-detuned near-resonant optical lattices}\label{sec:bluelatt}

The light shifts of the sublevels of the atomic ground state energy states are
positive for blue-detuned light. The stronger the coupling, the larger is the
light shift, and the higher the level lies in energy. Therefore the optical
potential minima may occur when the atomic ground state Zeeman sublevel is
minimally coupled to the excited states. This implies that it should be
possible to trap atoms with laser field polarization gradients around those
points in space where there is no coupling to the excited state. For example,
if the ground and excited states have an equal total angular momentum quantum
number $J$, the ground state sublevel with angular momentum projection state
$m_{g}=J$ is not coupled to the excited state when the light is $\sigma^{+}$
polarized. The coupling strength and the optical potential increases when the
atom moves away from the totally dark point, in this case away from the point
of $\sigma^{+}$ light polarization.

In the schemes for optical lattices which use e.g. only one ground and one
excited state manifold ($J\rightarrow J$), one of the light-shifted
eigenstates  is in fact a completely dark state which is not coupled to the
excited state at any point of space. It would be ideal for trapping atoms but
unfortunately the optical potential for this state is flat. This state can be
used for efficient laser cooling~\cite{Weidemuller94}, and the photon
scattering is about two orders of magnitude smaller than in other
near-resonant laser cooling and trapping schemes. In order to achieve a
lattice structure one can add a magnetic field~\cite{Grynberg94,Hemmerich95},
which  modifies the energy of the initially flat dark eigenstate in a periodic
way. Instead of applying a magnetic field one can also avoid the flat dark
state problem by using two different excited state hyperfine
manifolds~\cite{Stecher97}.

\begin{figure}[t!]
\noindent\centerline{\psfig{width=70mm,file=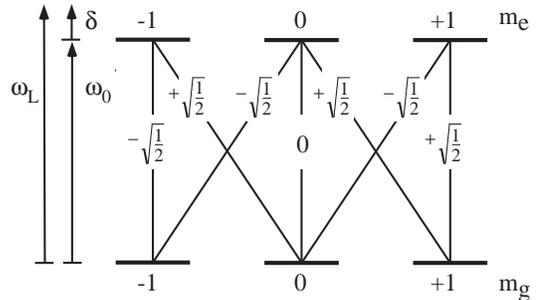} }
\caption[f1]{\label{fig:Levels}
The level structure of a single atom with the Clebsch--Gordan coefficients of
corresponding transitions. }
\end{figure}

In our study we consider the magneto--optical $J\rightarrow J$ lattice proposal
of Grynberg and Courtois (GC)~\cite{Grynberg94}. The numerical simulations for
dynamical collision studies in optical lattices are very
demanding~\cite{Piilo01a,Piilo01b}, and the GC scheme allows us to do
numerical simulations with parameter values that are also experimentally
realistic. We limit our study to the case where $J_{g}=J_{e}=1$. It is possible
to find a level scheme like this in $^{87}$Rb which has $F=1$ hyperfine states
for both the $5S_{1/2}$ ground and for the $5P_{1/2}$ excited
state~\cite{Hemmerich95}. We label the three ground state sublevels with
$|g_{\pm 1}>,|g_{0}>$, and the three excited state sublevels with $|e_{\pm 1}>,
|e_{0}>$, where the integer subscripts indicate the angular momentum
projection quantum number $m$, see Fig.~\ref{fig:Levels}. We use the atomic
mass of $^{87}$Rb in our simulations.

The laser field consists of two counter--propagating beams with orthogonal
linear polarizations and with frequency $\omega_L$. For a zero magnetic field
the laser detuning is $\delta=\omega_{L}-\omega_{0}$, where $\omega_{0}$ is
the atomic frequency. The total field has a polarization gradient in one
dimension and reads (with suitable choices of phases of the beams and origin
of the coordinate system)
\begin{equation}
        {\bf E}(z,t)={\cal E}_0 ({\bf e}_x e^{ik_rz} - i {\bf e}_y
        e^{-ik_rz})e^{-i\omega_L t} + c.c.,
        \label{eq:Efield}
\end{equation}
where ${\cal E}_0$ is the field amplitude and $k_r$ the wavenumber. With this
field, the polarization changes from circular $\sigma^-$ to linear and then to
circular in the opposite direction $\sigma^+$  when $z$ changes by
$\lambda_L/4$, where $\lambda_L$ is the wavelength of the laser light.

The intensity of the laser field and the strength of the coupling between the
field and the atom is described by  the Rabi frequency $\Omega = 2 d {\cal
E}_0 / \sqrt{2} \hbar$ where $d$ is the atomic dipole moment of the transition
between the ground and excited states. In the level scheme used here all the
allowed transitions are equally strong and the absolute value of the
corresponding Clebsch--Gordan coefficient has been included in the  definition
of the Rabi frequency $\Omega$ above.

The magnitude of the light shift is~\cite{Dalibard89} (when taking into account
the current level and coupling scheme)
\begin{equation}
       U_0=\frac{1}{2}\hbar \delta s_0, \label{eq:U_0}
\end{equation}
where $s_0$ is the saturation parameter given by
\begin{equation}
       s_0=\frac{\Omega^2/2}{\delta^2+\Gamma^2/4}. \label{eq:s_0}
\end{equation}

\begin{figure}[t!]
\noindent\centerline{\psfig{width=70mm,file=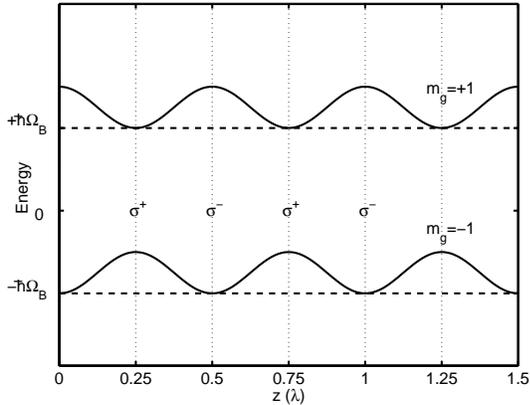} }
\caption[f2]{\label{fig:Lattice}
Schematic view of the optical potentials for the two trapping ground
state Zeeman sublevels. The periodic polarization gradient of the laser
field creates the lattice structure and we indicate the points of circular
polarizations $\sigma^{+}$ and $\sigma^{-}$. The dashed lines give the
Zeeman shifted energy levels which the light field modifies.}
\end{figure}

The magnetic field ${\bf B}$ is applied in the $z$--direction, $B=B_{z}$, which
is also the quantization axis. The ground state $m_{g}=0$ is not Zeeman shifted
but the states $m_{g}=\pm 1$ are shifted by $|\mu B| =\hbar \Omega_{B}$ into
opposite directions with respect to each other. Here $\mu$ is the magnetic
moment along the $z$--axis of the corresponding state. Thus the energy
separation between $m_{g}=\pm 1$ states due to magnetic field is $2
\hbar\Omega_{B}$.

If the light shift dominates over the Zeeman shift, $\hbar \Omega_{B} < U_{0}$,
the lattice is paramagnetic~\cite{Grynberg94}. The atoms are trapped in
potential wells located $0.5 \lambda$ apart, either in the points of
$\sigma^{+}$ polarization or in the points of $\sigma^{-}$ polarization,
depending on the direction  of the applied magnetic field. If the Zeeman shift
dominates, $U_{0} < \hbar \Omega_{B}$, the eigenstates of the system
correspond to Zeeman  sublevels onto which the light field induces
perturbations. In this case the atoms are trapped in both $\sigma^{-}$ and
$\sigma^{+}$ light polarization points in space and the lattice behaviour is
antiparamagnetic~\cite{Grynberg94}.

We have done our simulations in the antiparamagnetic regime where the Zeeman
shift dominates over light shifts, $U_{0}< \hbar \Omega_{B}$. The polarization
gradient of the laser field modifies the lattice potential as in the Sisyphus
scheme~\cite{Dalibard89}, see Fig.~\ref{fig:Lattice}. The atoms are cooled by
optical pumping, and are trapped into the $m_{g}=-1$ and $m_{g}=+1$ ground
sublevels. Without the magnetic field there would be no cooling  or trapping
for $J_{g}=J_{e}=1$ system due to the dark state and the lack of motional
coupling between the dark state and the coupled states~\cite{Jessen96}.

Since the Zeeman shifts for $m_{g}=-1$ and $m_{g}=+1$
are in opposite directions, their
effective detuning from the resonance is not equal and consequently they
experience different optical potential modulation depths $U_{-}$ and
$U_{+}$ respectively. The value of $U_{0}$ gives  the optical
potential modulation depth
when  the Zeeman shift is large compared to the detuning. This is the value
which both $U_{-}$ and $U_{+}$ approach when $U_{0}$ is kept fixed with
increasing detuning. We show the schematic view of optical potentials for
ground sublevels in Fig.~\ref{fig:Lattice}. The relevant lattice properties
along with simulation parameters are given in Tables~\ref{tab:Parameters1} and
\ref{tab:Parameters2}. We have performed a series of simulations for two fixed
values of detuning, with a changing Rabi frequency, and another series for a
fixed value of $U_{0}$, with a changing detuning and Rabi frequency.

\begin{table}[t]
\caption[t2]{\label{tab:Parameters1}
Laser parameters for the simulation series with fixed detuning, and the
corresponding lattice properties: Detuning $\delta$, Rabi frequency $\Omega$,
Zeeman shift $\Omega_{B}$, lattice modulation depths $U_0$, $U_{-}$, and
$U_{+}$. The atomic photon recoil energy is $E_{r} = \frac{\hbar^{2}
k_{r}^{2}} {2M}$, and $\Gamma$ is the atomic linewidth. }
\begin{tabular}{cccccccc}
         $\displaystyle{\delta} (\Gamma)$
       & $\displaystyle{\Omega} (\Gamma)$
       & $\displaystyle{\Omega_{B} (\Gamma)}$
       & $\displaystyle{U_0} (E_r)$
       & $\displaystyle{U_{-}} (E_r)$
       & $\displaystyle{U_{+}} (E_r)$\\
       \hline
      5.0   & 1.5  & 1.25   & 178  & 236  & 143  \\
      5.0   & 2.0  & 1.25   & 316  & 419  & 254  \\
      5.0   & 3.0  & 1.25   & 712  & 943  & 572  \\
      5.0   & 5.0  & 1.875  & 1980 & 3120 & 1447 \\
     10.0   & 2.0  & 1.25   & 159  & 182  & 141  \\
     10.0   & 4.22 & 1.875  & 710  & 873  & 599  \\
     10.0   & 8.0  & 3.125  & 2554 & 3704 & 1948 \\
     10.0   & 10.0 & 3.75   & 3990 & 6359 & 2905 \\
\end{tabular}
\end{table}

\begin{table}[t]
\caption[t2]{\label{tab:Parameters2}
Laser parameters for the simulation series with fixed lattice depth $U_0$, and
the corresponding lattice properties: Detuning $\delta$, Rabi frequency
$\Omega$, Zeeman shift $\Omega_{B}$, lattice modulation depths $U_0$, $U_{-}$,
and $U_{+}$. }
\begin{tabular}{cccccccc}

         $\displaystyle{\delta} (\Gamma)$
       & $\displaystyle{\Omega} (\Gamma)$
       & $\displaystyle{\Omega_{B} (\Gamma)}$
       & $\displaystyle{U_0} (E_r)$
       & $\displaystyle{U_{-}} (E_r)$
       & $\displaystyle{U_{+}} (E_r)$\\
       \hline
      1.5   & 1.72 & 0.8125 & 710  & 1126 & 489  \\
      5.0   & 3.0  & 1.25   & 712  & 943  & 572  \\
      7.0   & 3.54 & 1.25   & 712  & 865  & 605  \\
     10.0   & 4.22 & 1.875  & 710  & 873  & 599  \\
\end{tabular}
\end{table}

\section{Optical shielding} \label{sec:Shielding}

\subsection{Collisions in the present of near-resonant light}

Two slowly colliding atoms form a quasimolecule. When a laser field
is present the quasimolecule may be resonantly excited at long range at
the Condon point $R_{C}$ where the ground and excited molecular states
become resonant.  In this case the nature of the atomic collision depends
on the sign of the detuning of the laser~\cite{Weiner99}.

For a red-detuned laser the excitation is followed by an acceleration of the
quasimolecule on an attractive excited state. The collision becomes inelastic
when spontaneous decay back to the ground state occurs and the pair of
colliding atoms gain kinetic energy. The shared energy increase corresponds to
the acceleration on the excited state. This is the radiative heating
and escape mechanism. In addition the atoms may also gain kinetic energy by a
fine-structure changing mechanism. If the atom pair survives on the excited
state to very small internuclear distances and there is a level crossing
between two excited fine-structure states, the pair may change their internal
state. When coming out of the collision on a fine structure state which is
asymptotically below the state on which they entered the collision,
the kinetic energy of the quasimolecule increases by the corresponding
amount. For
a review see Ref.~\cite{Weiner99}.

\begin{figure}[t!]
\noindent\centerline{\psfig{width=70mm,file=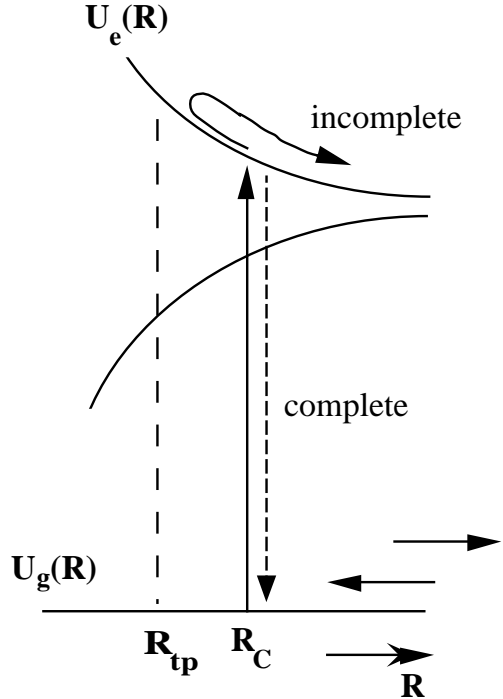} }
\caption[f3]{\label{fig:Shielding}
A schematic semiclassical representation of optical shielding. The
quasimolecule is excited resonantly to the repulsive molecular state at the
Condon point $R_{C}$. Then it reaches the classical turning point $R_{tp}$, and
is finally transferred back to the ground state when arriving at $R_{C}$ again.
If the transfer back to the ground state is not complete, the atom pair may
gain kinetic energy as it is further accelerated by the excited state
potential. In this case shielding is incomplete and the collision is
inelastic. If the population transfer between the states is adiabatic,
shielding is complete and the collision between the atoms is elastic. }
\end{figure}

When the laser field present is blue-detuned the situation is different. The
resonant excitation of the quasimolecule at $R_{C}$ occurs now to a
repulsive molecular state. The atom pair is prevented from approaching close
to each other due to the reflection of motion on the repulsive state at the
turning point $R_{tp}$, see Fig.~\ref{fig:Shielding}.  If the transfer of
population back to the ground state is not complete when the quasimolecule
traverses $R_{C}$ again the atom pair may gain kinetic energy. In this case
the shielding is incomplete and the collision is inelastic. If the population
transfer between the states is adiabatic and there is no time for spontaneous
emission to occur during the process, shielding becomes complete and the
laser-assisted collision between atoms is elastic.

It should be noted that despite strong localization of atoms into lattice
sites, atoms still move from site to site to some extent. This is because the
finite spatial extent of atomic wave packets and optical pumping process. Thus
the localization does not mean that weak long-distance atom-atom interactions
at relatively fixed atom-atom distances now dominate over the more
stronger cold collisions between moving atoms.

\subsection{Shielding of ground state processes}

Efficient optical shielding can suppress the rates of unwanted processes which
occur at short internuclear distances even in the absence of laser
fields~\cite{Weiner99}. Inelastic ground--ground processes are suppressed
because complete shielding simply empties the quasimolecule ground state at
$R_C$ and in principle the quasimolecule can never reach the internuclear
distances $R<R_C$. Since for near-resonant light the values for $R_C$ are
typically hundreds or thousands of {\AA}ngstroms, any small-distance molecular
process such as a change of a hyperfine state or Penning ionisation, is
prevented.

In the past the optical shielding and suppression has been studied
experimentally e.g. in the following contexts: suppression of heating and
escape of atoms in MOTs~\cite{Bali94}, shielding of
photoassociative ionizing collisions~\cite{Marcassa94}, shielding of
ionizing collisions of metastable xenon and
krypton~\cite{Katori94,Suominen96b}, and optical suppression in
two-photon ``energy pooling'' collisions in Rb MOTs~\cite{Sukenik98}.
Theoretical methods used for shielding studies include MCWF
simulations~\cite{Suominen95}, Landau--Zener~\cite{Suominen95}
and  three--dimensional Landau--Zener theory~\cite{Yurovsky97}, and quantum
close--coupling calculations~\cite{Napolitano97}. It should be noted that
shielding does not help very much in achieving Bose-Einstein
condensation as it
involves actual resonant absorption and subsequent emission of photons, with
the corresponding recoil momentum effects.

\subsection{Shielding in optical lattices}

When studying optical shielding in blue-detuned near-resonant optical lattices,
there is no need for additional shielding lasers. The blue-detuned lattice
laser itself acts as a shielding laser. Efficient shielding in blue-detuned
lattices would make it possible to increase the occupation density even in the
case of near-resonant lattices and without the problems arising from
light-assisted inelastic collisions and photon reabsorption. This would be in
contrast to the case of red-detuned near-resonant optical lattices where the
inelastic processes dominate when the occupation density increases and
atom--atom interactions are accounted~\cite{Piilo01a,Piilo01b}.

Our simulation scheme and collision model is similar to the one we have used
previously for collision studies in optical lattices and the details can be
found in Ref.~\cite{Piilo01b}.  The simulations require very large
computational resources and we note again that we fix the position of one
of the two colliding atoms in respect to the lattice. The small number of
scattered photons means that in a high density atom cloud the radiation
pressure due to reabsorbed photons  decreases. This means that our purely
collisional model becomes more realistic in the context of the thermodynamics
of the atom cloud, compared to our previous studies of the red-detuned
case~\cite{Piilo01a,Piilo01b}.

\section{Framework for simulations}\label{sec:Frame}

\subsection{Product state basis and Hamiltonian}

We formulate the problem in position space and use the two--atom product
states~\cite{Cohen-Tannoudji77}. We do not use the adiabatic elimination
of the excited states~\cite{Petsas99} since we want to account for the
dynamical nature of atomic interactions in the presence of
near-resonant light. The basis vectors are
\begin{equation}
       |j_{1} {m_{1}}\rangle_{1}
       |j_{2}  {m_{2}}\rangle_{2}, \label{eq:Basis}
\end{equation}
where $j_{1}$ and $j_{2}$ denote the ground or excited state  and
$m_{1}$, $m_{2}$ denote the quantum number for the component of $J$ along
the quantization axis $z$ for atom 1 and 2 respectively in our
$J_{g}=J_{e}=1$ system. The total number of states in principle is
$6\times 6=36$.  In the current coupling and level scheme it is possible
to simplify the problem and use only $9$ states, see below.

We have to fix the position of one atom, and the binary system wave function
depends now only on the position $z_{2}$ of the moving atom 2
\begin{equation}
       |\psi(z_2,t)\rangle  = \sum_{j_{1},j_{2},m_{1},m_{2}}
       \psi^{j_{1},m_{1}}_{j_{2},m_{2}}(z_2,t) |j_{1} {m_{1}}\rangle_{1}
       |j_{2} {m_{2}}\rangle_{2}. \label{eq:Psi}
\end{equation}
The atomic spatial dimensionality of the problem is reduced from numerically
impossible two to numerically treatable one, see Ref.~\cite{Piilo01b} for more
discussion. The relative coordinate $z$ between atoms is now $z=z_2-z_f$ where
$z_f$ is the position of the fixed atom.

In the atomic product state basis~\cite{Cohen-Tannoudji77}, our system
Hamiltonian is
\begin{equation}
       H_S=H_1+H_2+V_{dip} \label{eq:HS} .
\end{equation}
Here, $V_{dip}$ includes the interaction between the atoms. $H_1$ and $H_{2}$
are the single atom Hamiltonians for  atoms $1$ and $2$ respectively in
product state space.

The single atom Hamiltonian for atom $\alpha$ ($\alpha=1,2$) in respective
Hilbert spaces is, after the rotating wave approximation,
\begin{equation}
       \widetilde{H}_\alpha =  \frac{p_{\alpha}^{2}}{2M}
       - \hbar \delta P_{e,\alpha}
        + \widetilde{V}_\alpha
        +\widetilde{U}_{\alpha} \label{eq:HAlpha}.
\end{equation}
Here, $P_{e,\alpha} =\sum_{m=-1}^{1} |e_m \rangle_{\alpha}~_{\alpha}
\langle e_m|$, and the interaction between a single atom $\alpha$ and the
field is
\begin{eqnarray}
       \widetilde{V}_\alpha&=& i\frac{\hbar\Omega}{\sqrt{2}} \sin(kz_{\alpha})
       \left\{|e_{0} \rangle_{\alpha}~_{\alpha}\langle g_{-1}| +
       |e_{1} \rangle_{\alpha}~_{\alpha}\langle g_{0}|\right\} \nonumber \\
       && +\frac{\hbar\Omega}{\sqrt{2}}\cos(kz_{\alpha})
       \left\{|e_{-1} \rangle_{\alpha}~_{\alpha}\langle g_{0}| +
       |e_{0} \rangle_{\alpha}~_{\alpha}\langle g_{1}|\right\} +h.c.,
       \label{eq:VAtomLaser}
\end{eqnarray}
where $z_\alpha$ is the position operator of atom $\alpha$. The dipole--dipole
interaction potential $V_{dip}$ is the same as used in Refs.~\cite{Piilo01a}
and \cite{Piilo01b}. The sum over the quantum number $m$ in Eq.~(12) of
Ref.~\cite{Piilo01b} goes now from $m=-1$ to $m=+1$ for the level scheme
$J_{g}=J_{e}=1$ studied here.

The interaction of atom $\alpha$ with the magnetic field $B$
in Eq.~(\ref{eq:HAlpha}) is
\begin{equation}
     \widetilde{U}_{\alpha}=
     \sum_{i}m_i\hbar\Omega_{B_{i}} |i\rangle_{\alpha}~_{\alpha}\langle i|,
\end{equation}
where the sum over $i$ includes all the ground and excited states, and the
Zeeman shift factors $\Omega_{B_{i}}$ are for the ground substates $m=\pm1$
equal to
$\Omega_{B}$, see Tables~\ref{tab:Parameters1} and~\ref{tab:Parameters2}.

The description of optical shielding is usually given in terms of molecular
potentials, see Sec.~\ref{sec:Shielding}. We note that we use the two-atom
product state vectors due to their simplicity in treating quantums jumps for
the current case. Molecular potentials can be obtained by diagonalizing
$V_{dip}$ in Eq.~(\ref{eq:HS}).

In the antiparamagnetic regime of the GC optical lattice the values of
detunings are such that the Zeeman shifts are not negligible in all
simulations. The $m_{g}=-1$ level is closer to resonance than
the $m_{g}=+1$ level for the magnetic field orientation we use. Thus the
optical potential modulation depth may also be different for the two states.
The values of modulation depth $U_{-}$ and
$U_{+}$ for levels $m_{g}=-1$ and $m_{g}+1$ respectively and Zeeman
shifts are presented in
Tables~\ref{tab:Parameters1} and \ref{tab:Parameters2}. In the constant
detuning simulations it is necessary to increase the Zeeman shift when going
for stronger laser fields and deeper lattices, in order to stay in the
antiparamagnetic regime.

\subsection{Aspects of the simulation method}

We use the MCWF method to solve the equation
of motion for the density matrix of the two-atom system
interacting with semiclassical light
fields~\cite{Holland94,Garraway95,Suominen98}. A direct  quantum mechanical
solution of the density matrix master equation by other means is not feasible
due to the large size of the density matrix and the coupling of atoms to a
large number of vacuum modes of the electric field. The variant of Monte Carlo
(MC) methods we use was developed by Dalibard, Castin and
M{\o}lmer~\cite{Dalibard92}. Application of the MCWF method for the study of
cold collisions in optical lattices is not straightforward and we have given
the details in our earlier publication~\cite{Piilo01b}. As a result of
simulations we obtain a momentum distribution for an atom in the lattice. This
distribution is the most informative result when studying optical shielding in
the context presented here.

The laser field couples only states where $\Delta m=\pm 1$. The
Clebsch--Gordan coefficient between the states $m_{g}=0$ and $m_{e}=0$ is
zero in the level scheme. It follows that optical pumping rapidly moves
population to the ground levels $m_{g}=\pm 1$ which are coupled only
to the excited level $m_{e}=0$. Thus there is population on only three
levels for both of the colliding atoms, $\Lambda$ scheme, and number of
the used product state basis vectors can be reduced from 36  to $3\times
3=9$. This simplifies the problem considerably. The simulations take less
computational resources and are much faster to perform. The
dipole--dipole interaction does not change the scheme and it is confirmed
by the results of the full simulation which includes all 36 states. From
molecular potential point of view this means that only one attractive
and one repulsive state is relevant instead of all four different
attractive and four different repulsive states.

The number of collisions should be the same for simulations with different
parameters for the results to be comparable with each other. The number
of collisions is dictated by the number of spontaneously emitted photons.
We fix the simulation time to constant value 120, in units of $1/\Gamma
s_{0}$. This guarantees that the average number of spontaneously emitted
photons and the number of collisions remains roughly the same for all
simulations. We have performed all the simulations for the $1D$ occupation
density of $25\%$ of the lattice, i.e., every fourth lattice site is occupied
on average.

\section{Results}\label{sec:Results}

We present the momentum distributions of three simulation series in
Figs.~\ref{fig:KdistsD5}, \ref{fig:KdistsD10}, and \ref{fig:KdistsUo700}. We
do two simulation series with the constant values of detuning $\delta$ and
changing the Rabi frequency $\Omega$. We have $\delta=5\Gamma$ and
$\delta=10\Gamma$, in Figs.~\ref{fig:KdistsD5} and \ref{fig:KdistsD10},
respectively. Figure~\ref{fig:KdistsUo700} gives results for a constant optical
potential modulation depth, $U_{0}\sim710 E_{r}$, and changing $\delta$.
We compare the results between interacting and non--interacting atoms.

\subsection{Constant detuning}

\begin{figure}[t!]
\noindent\centerline{\psfig{width=70mm,file=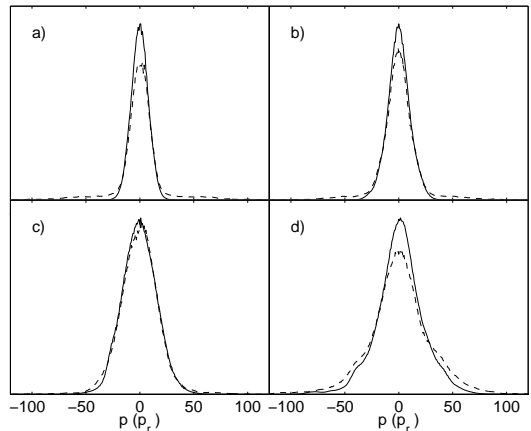} }
\caption[f2]{\label{fig:KdistsD5}
Momentum probability distributions for $\delta=5 \Gamma$ case. The Rabi
frequencies are: a) $\Omega=1.5\Gamma$, b) $\Omega=2.0\Gamma$, c)
$\Omega=3.0\Gamma$, and  d) $\Omega=5.0\Gamma$. Dashed line is for
interacting and solid line for non--interacting atoms. When $\Omega$
increases the regime changes from incomplete shielding, a) and b),
to complete shielding, c),  and finally to off--resonant heating in d).
The momentum distributions also get wider due to deeper lattice with
increasing $\Omega$. }
\end{figure}

\begin{figure}[t!]
\noindent\centerline{\psfig{width=70mm,file=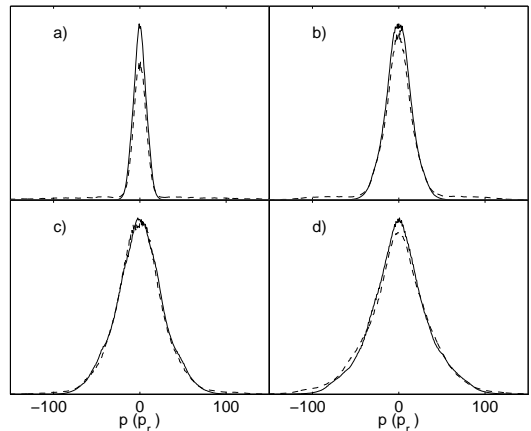} }
\caption[f2]{\label{fig:KdistsD10}
Momentum probability distributions for $\delta=10 \Gamma$ case. The Rabi
frequencies are: a) $\Omega=2.0\Gamma$, b) $\Omega=4.22\Gamma$, c)
$\Omega=8.0\Gamma$, and  d) $\Omega=10.0\Gamma$. Dashed line is for
interacting and solid line for non--interacting atoms. When $\Omega$
increases the regime changes from incomplete shielding, a) and b),
to complete shielding, c),  and finally to off--resonant effects in d).
The momentum distributions also get wider due to deeper lattice with
increasing $\Omega$. }
\end{figure}

For the smallest  values of $\Omega$, the constant detuning simulations
show that optical shielding is not complete. Wide wings towards large
momentum appear, Figs.~\ref{fig:KdistsD5}a) and b), \ref{fig:KdistsD10}a)
and b).  The resonant excitation process at Condon point $R_{C}$ is not
adiabatic and some atoms can move apart on repulsive molecular state after
the collision. If the process terminates for spontaneous decay at
$r>R_{C}$ the atom pair gains corresponding amount of kinetic energy by
acceleration on repulsive molecular state, see Fig.~\ref{fig:Shielding}.

The wings get narrower when we increase the value of $\Omega$ and nearly
vanish when the point of complete shielding is reached,
Figs.~\ref{fig:KdistsD5}c) and \ref{fig:KdistsD10}c).
The resonant excitation process for the repulsive state at $R_{C}$
becomes adiabatic and hardly any change in $p$--distribution is visible
when compared to non-interacting atoms. When atoms try to occupy the same
lattice site, they collide elastically in this parameter regime.

When we further increase the Rabi frequency, the off-resonant effects come
into play, Figs.~\ref{fig:KdistsD5}d) and~\ref{fig:KdistsD10}d). The
momentum distribution begin to deviate again from the result
for non--interacting atoms, and the qualitive character of momentum
broadening is now different when compared to the small $\Omega$ results.
This indicates that the heating process does not arise from incomplete
shielding.
Additional shoulders that appear in
both the interacting and non--interacting cases arise due to the
difference in the depth of the two lattice potential wells, $U_{\pm 1}$.
When we increase $\delta$, we note that the point of complete shielding
moves towards larger $U_{0}$. In our case from $\delta=5\Gamma$
and $U_{0}=712E_r$ to $\delta=10\Gamma$ and $U_{0}=2554E_r$.

\subsection{Constant lattice depth}

\begin{figure}[t!]
\noindent\centerline{\psfig{width=70mm,file=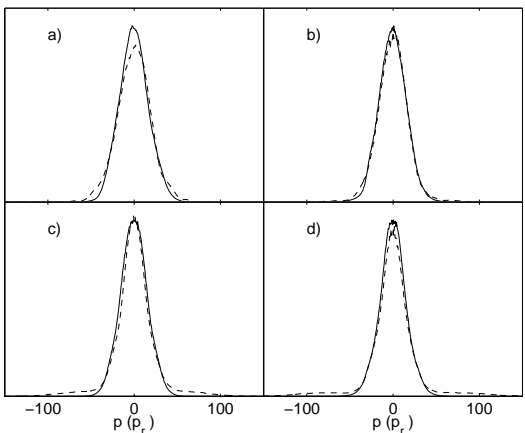} }
\caption[f2]{\label{fig:KdistsUo700}
Momentum distributions for fixed $U_{0}\sim 710E_{r}$.
a) $\delta=1.5\Gamma$
b) $\delta=5.0\Gamma$
c) $\delta=7.0\Gamma$
d) $\delta=10.0\Gamma$.
The off-resonant processes play a role at the small detuning, a). In the
intermediate detuning shielding may become complete and the collisions
between atoms are elastic, b). For larger detuning the shielding is
incomplete, c) and d).}
\end{figure}

Figure~\ref{fig:KdistsUo700} shows results for the case of constant optical
modulation depth $U_{0}\sim 710E_{r}$ and varying detuning. When the
detuning is small the off-resonant effects heat the atomic cloud and small
wings appear in momentum distribution, see Fig.~\ref{fig:KdistsUo700}a).
For
an intermediate detuning value the point of complete shielding is reached,
Fig.~\ref{fig:KdistsUo700}b). Increasing $\delta$ further makes the shielding
incomplete and subsequently wide wings appear in the momentum distribution,
Figs.~\ref{fig:KdistsUo700}c) and d). This is in agreement with the study of
off-resonant effects of the laser field in cold collisions
in MOTs~\cite{Suominen96a}. Keeping the lattice depth constant and increasing
the detuning means decreasing the ratio $\Omega/\delta$, which reduces
the role of off-resonant effects. Simulation studies reported in
Ref.~\cite{Suominen96a} show that for collisions in uniform blue-detuned
laser fields the off--resonant effects vanish when $\Omega/\delta$
decreases until for small values of $\Omega/\delta$ the shielding effect is
again incomplete. This  is accordance with the optical lattice results
presented here.

\section{Conclusions}\label{sec:Conclusions}

Our simulations demonstrate clearly that optical shielding should be present in
blue-detuned near-resonant optical lattices. The shielding process is not
expected to be always perfect but with an appropriate choice of the detuning
and the intensity of the trapping laser it is nevertheless possible to strongly
suppress the rate of inelastic collisions. For example, with a detuning
$\delta=5.0\Gamma$ and a Rabi frequency $\Omega=3.0\Gamma$, giving
$U_0=710E_r$ in our scheme, the optical shielding process is practically
complete, see Fig.~\ref{fig:KdistsD5}c). If the parameters are not chosen
correctly, the shielding process can be either incomplete or reduced by the
off-resonant excitation of the quasimolecule states related to attractive
atom-atom potentials~\cite{Suominen96a}.

We have studied here the antiparamagnetic regime of a
GC--lattice~\cite{Grynberg94}. In this case the cooling and trapping mechanism
resembles the traditional Sisyphus mechanism \cite{Dalibard89},
making comparisons between our
previous study of the red-detuned case more appropriate. In this scheme the
lattice depth depends on the intensity of the laser. The value of detuning for
effective shielding increases when the laser intensity and thus the optical
potential modulation depth $U_0$ is increased. Thus for shallow lattices
effective shielding occurs at small detunings, in which case the two trapping
potentials are not identical because of the non--negligible Zeeman shift
compared to the detuning of the laser.

In the paramagnetic regime of a GC--lattice the lattice depth $U_0$ is dictated
by the strength of the magnetic field more than by the laser intensity. This
leaves more freedom for using both the intensity and detuning of the laser to
optimize optical shielding for a given lattice depth.  This is an interesting
subject for possible future studies of collisions in blue-detuned
near-resonant optical lattices.

Together with the reduced photon scattering and thus reduced heating of dense
atomic clouds by photon reabsorption, optical shielding opens a way to achieve
higher occupation densities and lower temperatures in blue-detuned
near-resonant optical lattices, compared to the red-detuned case. Unlike in
the red-detuned case, two atoms trying to occupy the same lattice site will
at the worst case only repulse each other in an elastic process. It should be
noted that optical shielding was originally introduced to e.g. suppress the
hyperfine state changing ground-ground collisions during cooling and
trapping of alkali-metal atoms~\cite{Bali94}, and this additional benefit is
naturally present in blue-detuned lattices as well. More importantly, the
strong Penning ionisation for metastable rare gas
atoms~\cite{Katori94,Suominen96b} should be suppressed as well in blue-detuned
lattices. An important benefit here is that all three processes, trapping,
cooling and shielding, are produced by the same set of laser beams.

Considering what is already known about blue-detuned optical lattices, and
about optical shielding in magneto-optical traps, our results appear rather
obvious. However, they could have been different. Firstly, previous
theoretical studies of optical shielding have excluded the actual cooling
process~\cite{Suominen95,Yurovsky97,Napolitano97}. Our results show that the
two can coexist at least in blue-detuned lattices without apparent problems
even in the saturation limit, although the off-resonant excitation of
quasimolecule states reduces the shielding efficiency. In experiments in
magneto-optical traps shielding has been implemented typically with an
additional catalyst laser, with the red-detuned cooling and trapping lasers
either off or on. If the red-detuned cooling and trapping lasers are on, the
near-resonant shielding studies become practically impossible due to the
strong and usually unfavourable mixing of the two processes~\cite{Suominen96b}.
In other words, we hope that our work serves as a motivation for
experimentalists to perform shielding studies in near-resonant blue-detuned
lattices in the presence of the cooling process. Although the densities that
have been available so far have been rather low, the use of e.g. metastable
rare gas atoms could provide useful information due to the clear ion signal
that marks collision events~\cite{Katori94}.

Secondly, very few theoretical studies of shielding with near-resonant light
have been done so far, as the interest has been mostly on the large-detuned
laser fields. This is because the theoretical treatment is simplified when we
can ignore the spontaneous emission (at large detunings the excited state
potential is so steep that the resulting fast dynamics and slow spontaneous
emission decouple trivially from each other). In an earlier near-resonant
study~\cite{Suominen96a} the possibility of spontaneous emission from the
excited repulsive quasimolecule state during the slow-down of the relative
motion of the two atoms was found to be suppressed by re-excitation, until the
atoms had finally turned around, i.e., the efficiency of shielding increased
without limit with laser intensity. Similarly, in our study there does not
seem to be any evidence for breakdown of efficient shielding due to such a
spontaneous emission effect.

Thirdly, experiments show clearly that the shielding efficiency of inelastic
ground-ground collisions can saturate to some finite value as we increase
the laser intensity~\cite{Katori94,Suominen96b}. The reason for this is still
unclear. It has been attributed to various processes, in addition to the
above-mentioned premature termination of shielding via spontaneous
emission~\cite{Suominen95}. Other possibilities include counterintuitive or
off-resonant processes involving different partial waves or other processes
that similarly involve multiple states (in contrast to the basic two-state
approaches~\cite{Weiner99,Suominen96b,Yurovsky97,Napolitano97}), or
off-resonant excitation to the attractive molecular states~\cite{Katori94}. We
find that off-resonant excitation is indeed present in our studies, where
$\delta\leq 10\Gamma$, i.e., our detuning is small in general and comparable
to the Rabi frequencies. This is in agreement with the much more simplified
three-state study done earlier~\cite{Suominen96a}. When
$\Omega/\delta\gg 1$, the steady state formation can surpass the dynamical
resonant excitation process at suitably low collision velocities, and we see
that indeed it is the near-zero region of the momentum distribution that gets
depleted in Fig. 4(d), and not the near-average velocity region as in the
red-detuned lattice case. However, as discussed before, this off-resonant
process can not explain the observed saturation of shielding, as that happens
at very large detunings, for which $\Omega/\delta\ll 1$, but $\Omega$ is large
enough for the situation to correspond mainly to Fig. 4(c), the case of
complete shielding.  There is no evidence of saturation of shielding in our
studies, but due to the limitations of our model (one dimension only, and
ignoring the very small scale molecular processes), we can not state
conclusively that it would not be present in any experiments.

Thus, we have demonstrated that efficient optical shielding should be
possible in near-resonant blue-detuned optical lattices. More precisely, it is
mediated by the trapping and cooling lasers themselves but it does not
interfere with the cooling process, unlike in basic magneto-optical traps
created with strong red-detuned lasers. Especially, we have theoretically
demonstrated optical shielding within a level scheme which is more complicated
(and realistic) than the two-state models used so
far~\cite{Weiner99,Suominen95}. Our approach is fully quantum, takes
spontaneous emission into account, and is applicable even for strong fields.
Finally, our results show that by a suitable choice of lattice parameters it
should be possible to obtain efficient cooling and trapping, together with
efficient optical shielding of collisional processes, and yet to avoid any
contributions from off-resonant excitation to the attractive quasimolecule
states.

\acknowledgments

We thank Kirstine Berg-S{\o}rensen for helpful discussions. J.P. and
K.-A.S. acknowledge the Academy of Finland (projects 43336 and 50314) and the
European Union IHP CAUAC project for financial support, and the Finnish Center
for Scientific Computing (CSC) for computing resources. J.P. acknowledges
support from the National Graduate School on Modern Optics and Photonics.


\begin{references}

\bibitem{vanderStraten99} H. J. Metcalf and P. van der Straten,
	                       		{\it Laser Cooling and 
Trapping}, (Springer,
                            Berlin, 1999).


\bibitem{Jessen96}     P. S. Jessen and I. H. Deutsch,
                        Adv. At. Mol. Opt. Phys. {\bf 37}, 95 (1996);
                        D. R. Meacher,
                        Contemp. Phys.  {\bf 39}, 329 (1998);
                        S. Rolston,
                		      Phys. World {\bf 11} (10), 27 (1998);
                        L. Guidoni and P. Verkerk,
                		      J. Opt. B {\bf 1}, R23 (1999).

\bibitem{Dalibard89}  J. Dalibard and C. Cohen-Tannoudji,
		                    J. Opt. Soc. Am. B {\bf 6}, 2023 (1989);
                        P. J. Ungar, D. S. Weiss, E. Riis, and S. Chu,
		                    J. Opt. Soc. Am. B {\bf 6}, 2058 (1989).

\bibitem{Hensinger01} W. K. Hensinger, H. H\"affner, A. Browaeys,
                        N. R. Heckenberg, K. Helmerson, C. McKenzie,
                        G. J. Milburn, W. D. Phillips, S. L. Rolston,
                        H. Rubinsztein-Dunlop, and B. Upcroft,
                        Nature {\bf 412}, 52 (2001).

\bibitem{Greiner02}   M. Greiner, O. Mandel, T. Esslinger,
                        T. W. H\"ansch, and I. Bloch,
                        Nature {\bf 415}, 39 (2002).

\bibitem{Pedri01}     S. Burger, F. S. Cataliotti, C. Fort, F. Minardi,
                       M. Inguscio,  M. L. Chiofalo, and M. P. Tosi,
                       Phys. Rev. Lett. {\bf 86}, 4447 (2001);
                       P. Pedri, L. Pitaevskii, S. Stringari,
                       C. Fort, S. Burger, F. S. Cataliotti,
                       P. Maddaloni, F. Minardi, and M. Inguscio,
                       Phys. Rev. Lett. {\bf 87}, 220401 (2001).

\bibitem{Morsch01}    O. Morsch, J. H. M\"uller, M. Cristiani, D. Ciampini,
                       and E. Arimondo,
                       Phys. Rev. Lett. {\bf 87}, 140402 (2001).


\bibitem{Holland94}    M. J. Holland, K.-A. Suominen, and K. Burnett,
                        Phys. Rev. Lett. {\bf 72}, 2367 (1994);
                        Phys. Rev. A {\bf 50}, 1513 (1994).

\bibitem{Weiner99}     K.-A. Suominen,
                        J. Phys. B {\bf 29}, 5981 (1996);
                        J. Weiner, V. S. Bagnato, S. Zilio, and P. S. Julienne,
                        Rev. Mod. Phys. {\bf 71}, 1 (1999) and
                        references therein.

\bibitem{Jaksch98}    D. Jaksch, C. Bruder, J. I. Cirac, C. W. Gardiner,
                       and P. Zoller,
		                    Phys. Rev. Lett. {\bf 81}, 3108 (1998).

\bibitem{Brennen99}    G. K. Brennen, C. M. Caves, P. S.
                        Jessen, and I. H. Deutsch,
		                     Phys. Rev. Lett. {\bf 82}, 1060 (1999);
                        G. K. Brennen, I. H. Deutsch, and P. S. Jessen,
                        Phys. Rev. A {\bf 61}, 062309 (2000);
                        I. H. Deutsch, G. K. Brennen, and P. S. Jessen,
                        Fortschr. Phys. {\bf 48}, 925 (2000).

\bibitem{Jaksch99}  D. Jaksch, H.-J. Briegel, J. I. Cirac, C. W. Gardiner,
                     and P. Zoller,
		                  Phys. Rev. Lett. {\bf 82}, 1975 (1999);
                     H.-J. Briegel, T. Calarco, D. Jaksch, J. I. Cirac, and
                     P. Zoller,
                     J. Mod. Opt. {\bf 47}, 415 (2000).

\bibitem{DePue99}      M. T. DePue, C. McCormick, S. L. Winoto, S. Oliver, and
                         D. S. Weis,
		                     Phys. Rev. Lett. {\bf 82}, 2262 (1999);
                         A. J. Kerman, V. Vuleti\'c, C. Chin, and S. Chu,
                         Phys. Rev. Lett. {\bf 84}, 439 (2000).

\bibitem{Grynberg94}  G. Grynberg and J.-Y. Courtois,
                        Europhys. Lett. {\bf 27}, 41 (1994);
                        K. I. Petsas, J.-Y. Courtois, and G. Grynberg,
                        Phys. Rev. A {\bf 53}, 2533 (1996).

\bibitem{Hemmerich95}   A. Hemmerich, M. Weidem{\"u}ller, T. Esslinger,
                          C. Zimmermann, and T. H{\"a}nsch,
                          Phys. Rev. Lett. {\bf 75}, 37 (1995).

\bibitem{Bali94}    S. Bali, D. Hoffmann, and T. Walker,
                      Europhys. Lett. {\bf 27}, 273 (1994);
                      V. Sanchez-Villicana, S. D. Gensemer, K. Y. N. Tan,
                      A. Kumarakrishnan, T. P. Dinneen, W. S{\"u}ptitz, and
                      P. L. Gould,
                      Phys. Rev. Lett. {\bf 74}, 4619 (1995);
                      S. R. Muniz, L. G. Marcassa, R. Napolitano, G. D. Telles,
                      J. Weiner, S. C. Zilio, and V. S. Bagnato,
       	              Phys. Rev. A {\bf 55}, 4407 (1997).

\bibitem{Marcassa94}  L. Marcassa, S. Muniz, E. de Queiroz, S. Zilio,
                        V. Bagnato,
                        J. Weiner, P. S. Julienne, and K.-A. Suominen,
                        Phys. Rev. Lett. {\bf 73}, 1911 (1994).

\bibitem{Katori94}  H. Katori and F. Shimizu,
                       Phys. Rev. Lett. {\bf 73}, 2555 (1994);
                       M. Walhout, U. Sterr, C. Orzel, M. Hoogerland,
                      and S. L. Rolston,
                        Phys. Rev. Lett. {\bf 74}, 506 (1995).

\bibitem{Suominen96b} K.-A. Suominen, K. Burnett, P. S. Julienne,
                        M. Walhout, U. Sterr, C. Orzel, M. Hoogerland,
                        and S. L. Rolston,
                        Phys. Rev. A {\bf 53}, 1678 (1996).

\bibitem{Sukenik98} C. I. Sukenik, D. Hoffmann, S. Bali, and T. Walker,
		                  Phys. Rev. Lett. {\bf 81}, 782 (1998).

\bibitem{Suominen95}  K.-A. Suominen, M. J. Holland, K. Burnett,
                        and P. S. Julienne,
                        Phys. Rev. A {\bf 51}, 1446 (1995).

\bibitem{Yurovsky97} V. A. Yurovsky and A. Ben-Reuven,
		                   Phys. Rev. A {\bf 55}, 3772 (1997).

\bibitem{Napolitano97} R. Napolitano, J. Weiner, and P. S. Julienne,
                         Phys. Rev. A {\bf 55}, 1191 (1997).

\bibitem{Piilo01a}    J. Piilo, K.-A. Suominen, and K. Berg-S{\o}rensen,
                        J. Phys. B {\bf 34}, L231 (2001).

\bibitem{Piilo01b}    J. Piilo, K.-A. Suominen, and K. Berg-S{\o}rensen,
                       to appear in Phys. Rev. A March 2002.

\bibitem{Weidemuller94} M. Weidem{\"u}ller, T. Esslinger,
                          M. A. Ol'Shanii, A. Hemmerich, and	T. H{\"a}nsch,
			                     Europhys. Lett. {\bf 
27}, 109 (1994).

\bibitem{Stecher97}   H. Stecher, H. Ritsch, P. Zoller, F. Sander,
                        T. Esslinger, and T. W. H{\"a}nsch,
		                    Phys. Rev. A {\bf 55}, 545 (1997).

\bibitem{Cohen-Tannoudji77} C. Cohen--Tannoudji, B. Diu, and F. Lalo\"e,
                              {\it Quantum Mechanics} Vol. I
                              (Wiley--Interscience, Paris, 1977),
                              Chapter II F.


\bibitem{Petsas99}    K. I. Petsas, G. Grynberg, and J.-Y. Courtois,
		                    Eur. Phys. J. D {\bf 6}, 29 (1999) and
                        references therein.

\bibitem{Garraway95}    B. Garraway and K.-A. Suominen,
                          Rep. Prog. Phys. {\bf 58}, 365 (1995)
                          and references therein.


\bibitem{Suominen98}  K.-A. Suominen, Y. B. Band, I. Tuvi, K. Burnett,
                        and P. S. Julienne,
                        Phys. Rev. A {\bf 57}, 3724 (1998).

\bibitem{Dalibard92}  J. Dalibard, Y. Castin, and K. M{\o}lmer,
                          Phys. Rev. Lett. {\bf 68}, 580 (1992);
                          K. M{\o}lmer, Y. Castin, and J. Dalibard,
                          J. Opt. Soc. Am. B {\bf 10}, 524 (1993).

\bibitem{Suominen96a} K.-A. Suominen, K. Burnett, and P. S. Julienne,
                        Phys. Rev. A {\bf 53}, R1220 (1996).



\end{references}
\end{document}